\documentclass[12pt]{iopart}

\usepackage{graphicx}
\usepackage{iopams}

\newcommand{\mb}{\mathbf}

\begin{document}

\title[Physical interpretation of the Wigner rotations and relativistic quantum information]{Physical interpretation of the Wigner rotations and its implications for relativistic quantum information }

\author{Pablo L Saldanha$^{1,2}$ and Vlatko Vedral$^{1,3,4}$}
\address{$^1$ Department of Physics, University of Oxford, Clarendon Laboratory, Oxford, OX1 3PU, United Kingdom}
\address{$^2$ Departamento de F\'isica, Universidade Federal de Pernambuco, 50670-901, Recife, PE, Brazil}
\address{$^3$ Centre for Quantum Technologies, National University of Singapore, Singapore}
\address{$^4$ Department of Physics, National University of Singapore, Singapore}

\ead{saldanha@df.ufpe.br}

\begin{abstract}
We present a new treatment for the spin of a massive relativistic particle in the context of quantum information based on a physical interpretation of the Wigner rotations, obtaining different results in relation to the previous works. We are lead to the conclusions that it is not possible to define a reduced density matrix for the particle spin and that the Pauli-Lubanski (or similar) spin operators are not suitable to describe measurements where spin couples to an electromagnetic field in the measuring apparatus. These conclusions contradict the assumptions made by most of the previous papers on the subject. We also propose an experimental test of our formulation.
\end{abstract}



\maketitle

\section{Introduction}

The field of relativistic quantum information has recently emerged \cite{czachor97,peres02,alsing02,gingrich02,ahn03,terno03,terashima03,czachor03,li03,peres04,lee04,bartled05,kim05,czachor05,peres05,caban05,jordan06,lamata06,jordan07,landulfo09,dunningham09,caban10,friis10,choi11}, describing how relativistic particles behave in a regime where the nature and the number of particles do not change in the processes, such that not all the machinery of quantum field theory is necessary. This simplified view of the problems can shed light on many issues of relativistic quantum mechanics and may have applications in the near future if we use the spin of  relativistic particles to encode quantum information. Latter the field has expanded to include non-inertial reference frames and general relativity effects, where the number of particles may not be conserved \cite{alsing03,fuentes05,alsing06,fuentes10,palmer11}.

Here we present a new treatment for the problem based on a physical interpretation of the Wigner rotations \cite{wigner39,weinberg}, which specify a momentum-dependent change of the spin state of a particle with a change of reference frame. We show that the Wigner rotations are consistent with the fact that different observers compute different quantization axes for a spin measurement, being a direct consequence of the dependence of the  quantization axis of a spin measurement on the particle momentum. We are lead to the conclusion that it is not possible to make a momentum-spin separation of the system and to define a reduced density matrix for spin, as is done in many previous papers on the subject \cite{peres02,gingrich02,li03,peres04,bartled05,peres05,caban05,jordan06,lamata06,jordan07,landulfo09,dunningham09,friis10,choi11}. We also show that the use of the Pauli-Lubanski (or similar) spin operators to describe spin measurements, as in  \cite{czachor97,ahn03,czachor03,lee04,kim05,czachor05,caban05,caban10,friis10}, depends on the coupling of the spin to a quantity that transforms as part of a 4-vector under the Lorentz transformations in the measuring apparatus. However, we do not know if such a coupling exists in nature. Our treatment assumes that spin couples to the electromagnetic field in the measuring apparatus, as in the Stern-Gerlach experiment, and it consequently makes different predictions for the expectation values of spin measurements in relation to the previous treatments. We also propose here an experimental test of our formulation.

\section{Physical interpretation of the Wigner rotations}

Weinberg's treatment \cite{weinberg} for the Wigner rotations for massive particles is reproduced in the Appendix. The conclusion is that representing the quantum state of a relativistic particle with mass $m$, 4-momentum $p=(p^0,\mb{p})$ and spin state $\phi$ as $|{p,\phi}\rangle$, with a change of reference frame represented by a homogeneous Lorentz transformation $\Lambda$, the particle state in the new frame is \cite{weinberg}
\begin{equation}\label{wigner}
    U(\Lambda)|{p,\phi}\rangle=\sqrt{\frac{(\Lambda p)^0}{p^0}}\sum_{\phi'}D_{\phi,\phi'}(W)|{\Lambda p,\phi'}\rangle,
\end{equation}
where $U(\Lambda)$ is  the corresponding unitary transformation  and $\sqrt{(\Lambda p)^0/{p^0}}$ is a normalization factor. The particle 4-momentum in the new frame is $\Lambda p$. The Wigner rotation $W(\Lambda,p)\equiv L^{-1}(\Lambda p)\Lambda L(p)$, where $L(p)$ represents a Lorentz boost that transforms the 4-momentum  $(m,0)$ in the 4-momentum $p=(p^0,\mb{p})$ and $L^{-1}$ is the inverse of $L$, changes the particle spin state via the matrix $D_{\phi,\phi'}(W)$. In this paper we are using a system of units in which the speed of light in vacuum is $c=1$.

We would now like to present a physical interpretation of the change of the particle spin state with the change of reference frame indicated by  (\ref{wigner}). The particle spin is defined to be the angular momentum it has in its own rest frame. Let us consider a spin-1/2 particle. In the particle rest frame, where the 4-momentum is $(m,0)$ and a non-relativistic treatment can be used with a momentum-spin separation, the particle spin state can be described by the Bloch vector $\mb{r}\equiv \langle \phi|\hat{{\bsigma}}|\phi\rangle$, that represents the expectation value of the Pauli matrices $\hat{{\bsigma}}\equiv\hat{\sigma}_x\mb{\hat{x}}+\hat{\sigma}_y\mb{\hat{y}}+\hat{\sigma}_z\mb{\hat{z}}$. If we substitute  in  (\ref{wigner}) the labels $\phi$ by the labels $\mb{r}$, that contain the same amount of information for a spin-1/2 particle, the Wigner rotation will change the Bloch vector as a 3-dimensional rotation $\mb{r}'=R(W)\mb{r}$. So  (\ref{wigner}) can be written as
\begin{equation}\label{wigner_r}
    U(\Lambda)|{p,\mb{r}}\rangle=\sqrt{\frac{(\Lambda p)^0}{p^0}}|{\Lambda p,R(W)\mb{r}}\rangle.
\end{equation}

The treatment so far is standard, but now we ask the crucial question: how can we prepare the state $|{p,\mb{r}}\rangle$? According to quantum mechanics \cite{peres}, we have to measure the particle momentum, obtaining eigenvalues $p^i$ for each component, and measure spin with a quantization axis in the direction $\mb{r}$ in the particle rest frame obtaining an eigenvalue $+\hbar/2$ for the spin component. To measure the particle spin, we can use a Stern-Gerlach apparatus with a inhomogeneous magnetic field that points in the direction $\mb{r}$ in the particle rest frame. To find the magnetic field in the particle rest frame, we must apply the corresponding Lorentz transformation to the electromagnetic tensor of the apparatus field in the laboratory frame \cite{jackson}: $F^{(0)}=L^{-1}(p)F\tilde{L}^{-1}(p)$, where $\tilde{L}$ represents the transpose of $L$. Defining $\mb{b}(F)$ as an unitary vector in the direction of the magnetic field of the electromagnetic tensor $F$, we have
\begin{equation}\label{rbp}
    \mb{r}=\mb{b}(L^{-1}(p)F\tilde{L}^{-1}(p)).
\end{equation}

Now let us consider the description of the process of preparation of the particle state by an observer in a reference frame obtained from a homogeneous Lorentz transformation $\Lambda$ acting on the previous frame. The electromagnetic tensor of the Stern-Gerlach apparatus in the new frame is $\Lambda F\tilde{\Lambda}$ and the particle 4-momentum is $\Lambda p$. So the Bloch vector $\mb{r}'$ in the direction of the magnetic field  in the particle rest frame computed by this observer is $\mb{r}'=\mb{b}(L^{-1}(\Lambda p)\Lambda F\tilde{\Lambda}\tilde{L}^{-1}(\Lambda p))$. But the Wigner rotation is defined as $W(\Lambda,p)\equiv L^{-1}(\Lambda p)\Lambda L(p)$ and if we multiply both sides by $L^{-1}(p)$ from the right we obtain $WL^{-1}(p)=L^{-1}(\Lambda p)\Lambda$. So we have
\begin{equation}
    \mb{r}'=\mb{b}(WL^{-1}(p) F\tilde{L}^{-1}(p)\tilde{W})=R(W)\mb{b}(L^{-1}(p)F\tilde{L}^{-1}(p))=R(W)\mb{r},
\end{equation}
in agreement with  (\ref{wigner_r}). The Wigner rotation is a direct consequence of the dependence of the quantization axis of a spin measurement with the particle momentum, representing the fact that different observers compute different quantization axes for the measurement. For particles with spin higher than 1/2, the Bloch vector cannot be used to represent the spin state, but the physical interpretation of the Wigner rotation is the same.

\section{Expectation values of spin measurements on relativistic particles}

According to our formalism, the expectation value of the measurement of a spin component of a spin-1/2 particle prepared in the sate $|{p,\mb{r}}\rangle$ made by a Stern-Gerlach type measurement with electromagnetic field tensor $F$, considering eigenvalue $+1$ $(-1)$ if the spin is found aligned (anti-aligned) with the magnetic field in the particle rest frame, is
\begin{equation}\label{exp_val}
    E(|{p,\mb{r}}\rangle,F)=\mb{r}\cdot \mb{b}(L^{-1}(p)F\tilde{L}^{-1}(p)).
\end{equation}
This occurs because in the particle rest frame we can use non-relativistic quantum mechanics to state that the expectation value of the measurement is the scalar product between the Bloch vector and an unitary vector in the direction of the magnetic field in this frame. In a recent work, Palmer \textit{et al.} treated Stern-Gerlach measurements on relativistic particles that are in momentum eigenstates obtaining equivalent results \cite{palmer11}. We see that the expectation value of the spin measurement depends explicitly on the particle momentum. Since it is not possible to measure the spin of a relativistic particle in an independent way from its momentum, a spin-momentum partition of the system is meaningless. As a consequence, it is not possible to define a reduced spin matrix for the system tracing out the momenta, as it was done in many previous treatments of the subject
\cite{peres02,gingrich02,li03,peres04,bartled05,peres05,caban05,jordan06,lamata06,jordan07,landulfo09,friis10,choi11}, since it is not possible to correctly predict the outcome statistics of a spin measurement without considering the particle momentum. Spin and momentum cannot be treated as independent variables if the particle has relativistic speeds. 

To illustrate the impossibility of the definition of a reduced density matrix for a relativistic particle, consider the following example. A spin-1/2 particle in a superposition of different momenta is deflected up by a Stern-Gerlach apparatus. Since each momentum component have a different quantization axis for the measurement, there will be a correlation between the particle momenta and the particle spin after the measurement. So, if we trace out the particle momenta, we obtain a mixed reduced density matrix for the particle spin. Let us consider that the deflection of the particle is compensated by the application of electromagnetic fields such that the particle momentum distribution after the measurement is the same as before. If the particle spin is now measured by an identical apparatus, since each momentum component will have the same quantization axis as in the preparation procedure, we can state with 100\% certainty that the particle will be deflected up. But this characterizes a pure state of spin. The definition of a reduced density matrix for the spin of a relativistic particle leads to unavoidable paradoxes. 

For consistency of the treatment, we must be able to describe the interaction energy between the particle spin and the Stern-Gerlach apparatus electromagnetic field in a covariant way. This would guarantee that observers in any inertial reference frame predict the same expectation values for the measurements. This is important because if one observer registers a detection of a particle in one particular detector, then all other observers must agree that the same detector has registered that particle. When the experiment is repeated many times and the data accumulated, all observers must agree with the outcomes statistics. In the particle rest frame, the interaction Hamiltonian is $\hat{H}_{SG}=-\alpha \hat{\mb{s}}\cdot\mb{B}_0$, where $\mb{B}_0$ is the magnetic field in the particle rest frame, $\hat{\mb{s}}=\hbar\hat{{\bsigma}}/2$ is the particle spin operator and $\alpha$ is the gyromagnetic ratio, $\hat{{\bmu}}_0=\alpha \hat{\mb{s}}$ being the particle magnetic dipole moment operator in the rest frame. The magnetic dipole moment ${\bmu}$ and the electric dipole moment $\mb{d}$ of a particle multiplied by $\gamma_v\equiv1/\sqrt{1-v^2}$, where $\mb{v}$ is the particle velocity and $v\equiv |\mb{v}|$, form an anti-symmetric tensor $D$ in the same way as the electromagnetic tensor $F$, with the substitutions $\mb{E}\rightarrow\gamma_v\mb{d}$ and $\mb{B}\rightarrow-\gamma_v{\bmu}$ \cite{penfield}. Since in the particle rest frame the magnetic and electric dipole moment operators are $\hat{{\bmu}}_0=\alpha \hat{\mb{s}}$ and $\hat{\mb{d}}_0=0$, in a reference frame where the particle has velocity $\mb{v}$ the interaction Hamiltonian can be written as
\begin{eqnarray}\label{U}
    &&\hat{H}_{\mathrm{SG}}(p)=-\hat{{\bmu}}\cdot \mathbf{B}-\hat{\mb{d}}\cdot \mathbf{E}=-\frac{1}{2\gamma_v}\mathrm{Tr}(gFg\hat{D})\;,\\\nonumber
    &&\mathrm{with}\;\;\hat{{\bmu}}=\alpha\left[\hat{\mb{s}}-\frac{\gamma_v}{\gamma_v+1}\mb{v}(\mb{v}\cdot\hat{\mb{s}})\right]\;\;\mathrm{and}\;\;
    \hat{\mb{d}}=\alpha(\mb{v}\times\hat{\mb{s}}),
\end{eqnarray}
where $g$ is the diagonal matrix with elements $(1,-1,-1,-1)$ and $\mathrm{Tr}$ stands for the trace.  $\hat{H}_{\mathrm{SG}}$ depends on the 4-momentum $p$, since the operators $\hat{{\bmu}}$ and $\hat{\mb{d}}$ depend on the particle velocity. The treatment is covariant because if we make a change of reference frame represented by a Lorentz transformation $\Lambda$, the interaction Hamiltonian in the new frame obeys
\begin{equation}
    \gamma_{v'}\hat{H}_{\mathrm{SG}}'=-\frac{1}{2}\mathrm{Tr}(g\Lambda F\tilde{\Lambda}g\Lambda\hat{D}\tilde{\Lambda})=\gamma_{v}\hat{H}_{\mathrm{SG}},
\end{equation}
since $\tilde{\Lambda} g \Lambda=g$. We see that the spin operators encoded in the magnetic dipole moment operators must transform as part of a tensor under Lorentz transformations for the interaction to be described in a covariant way. The expectation value of spin measurements can then be written as $-\langle\hat{H}_{\mathrm{SG}}(p)\rangle/|\lambda(\hat{H}_{\mathrm{SG}}(p))|$ for each momentum component, being the same in all inertial reference frames, where $\langle\hat{H}_{\mathrm{SG}}(p)\rangle$ represents the expectation value of the interaction Hamiltonian and $|\lambda(\hat{H}_{\mathrm{SG}}(p))|$ the modulus of its eigenvalues. So we can write the operator related to a Stern-Gerlach spin measurement as
\begin{equation}\label{m_sg}
    \hat{M}_{\mathrm{SG}}=\int d^3p\,|p\rangle\langle p|\otimes \frac{\hat{H}_{\mathrm{SG}}(p)}{|\lambda(\hat{H}_{\mathrm{SG}}(p))|}
\end{equation}
with a representation of states in the basis $|p\rangle\otimes|\phi(p)\rangle$ that, despite the notation, cannot have a momentum-spin separation for the reasons described before. The expectation value of the spin measurement can then be written as $E=\mathrm{Tr(\hat{M}_{\mathrm{SG}}\rho)}$, where $\rho$ represents the particle quantum state, being the same in all inertial reference frames.

In the relativistic quantum information literature, many authors use the Pauli-Lubanski (or similar) spin operators to describe spin measurements on relativistic particles \cite{czachor97,ahn03,czachor03,lee04,kim05,czachor05,caban05,caban10,friis10}. Their description is mathematically covariant, in the sense that the expectation values they obtain for the measurements are the same in all inertial reference frames \cite{lee04}, but none of the authors describe a physical system capable of performing the measurements. In a physical implementation, the operator related to the measurement must be written in terms of the interaction Hamiltonian between the spin and the measuring apparatus, like in  (8). The Pauli-Lubanski spin operator $\hat{\mb{S}}$ is part of a 4-vector $\hat{W}=(\hat{W}^0,\hat{\mb{W}})\equiv(\hat{\mb{S}}\cdot\mb{p},p^0\hat{\mb{S}})$, where $p^0$ is the energy and $\mb{p}$ the momentum of the particle \cite{muirhead}. If we want to give a covariant description for the interaction of the Pauli-Lubanski spin with a measuring apparatus, such that the expectation value of a spin measurement be the same in all inertial reference frames, $\hat{W}$  must couple to a  4-vector quantity $(G^0,\mb{G})$ with an interaction Hamiltonian of the form $\hat{H}_\mathrm{PL}\propto \hat{W}^0G^0-\hat{\mb{W}}\cdot\mb{G}$. However, we don't know if such a coupling exists, and the physical implementation of the measurement depends on the existence of such coupling in nature. So the use of the Pauli-Lubanski (or similar) spin operators to describe measurements in  \cite{czachor97,ahn03,czachor03,lee04,kim05,czachor05,caban05,caban10,friis10} may have consistency problems, and certainly is not suitable to describe measurements where spin couples to the electromagnetic field.

\section{Experimental proposal to test our formulation}

\begin{figure}\begin{center}
  \includegraphics[width=8cm]{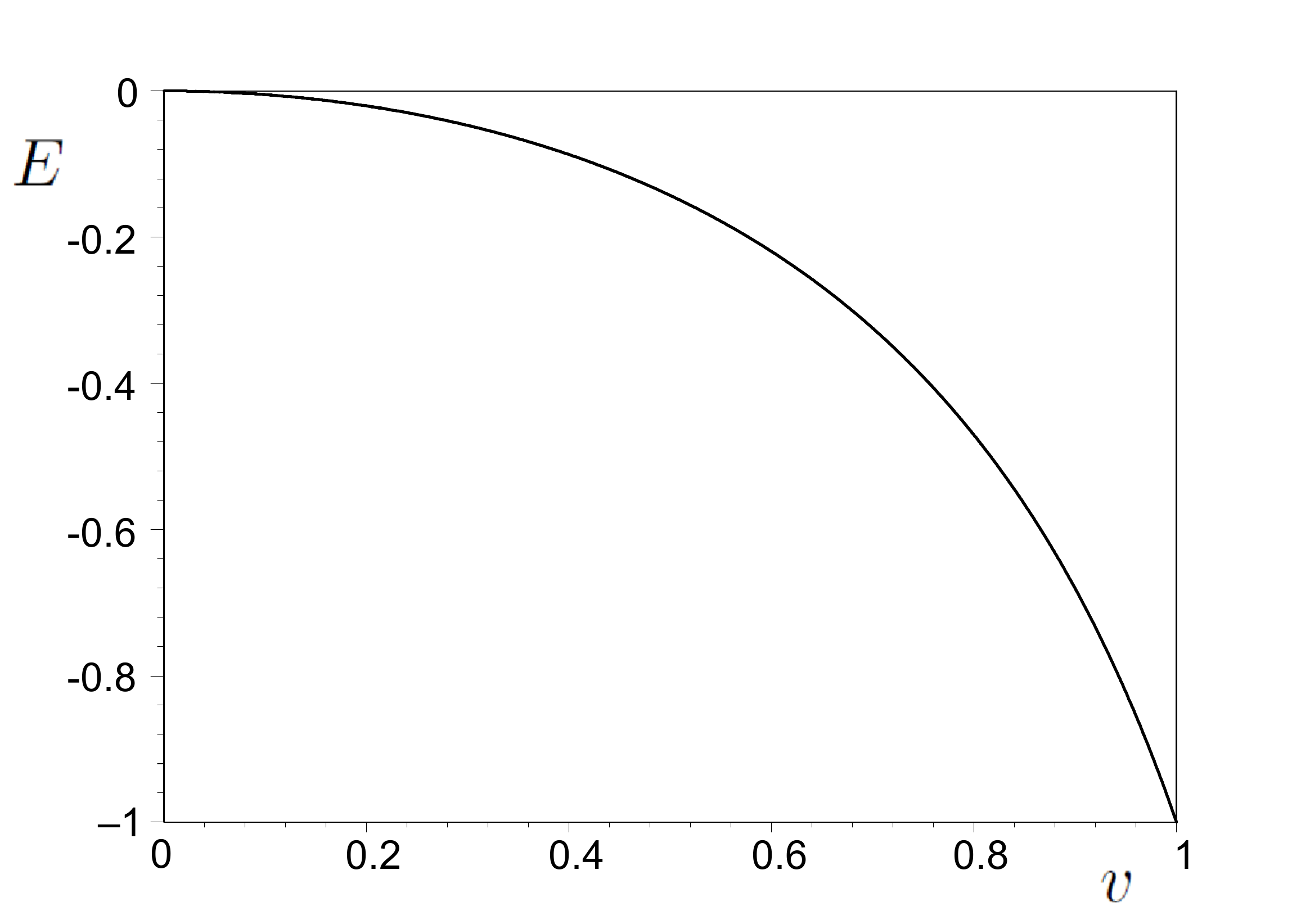}\\
  \caption{Expectation value for a spin measurement on a particle with velocity $v(\mb{\hat{x}}+\mb{\hat{y}})/\sqrt{2}$ that is prepared with eigenvalue $+1$ by a Stern-Gerlach apparatus with magnetic field in the $\mb{\hat{x}}$ direction made by a second apparatus with magnetic field that points in the $\mb{\hat{y}}$ direction.
}\label{fig1}
 \end{center}\end{figure}

As a possible experimental test of our formulation, let us consider a neutral spin-1/2 particle that propagates with velocity $\mathbf{v}=v[\cos(\theta)\mb{\hat{x}}+\sin(\theta)\mb{\hat{y}}]$, having momentum $\mb{p}=m\mb{v}/\sqrt{1-v^2}$, and passes through two Stern-Gerlach apparatuses, the first one with an inhomogeneous magnetic field in the $\mb{\hat{x}}$ direction and the second one with an inhomogeneous magnetic field in the $\mb{\hat{y}}$ direction.  Let us now compute the expectation value of the measurement of the second apparatus $E$ after the first one had yielded an eigenvalue $+1$ for the spin component. Using  (\ref{rbp}) to find the spin state prepared by the first apparatus and  (\ref{exp_val}) to find the expectation value of the measurement of the second apparatus we obtain
\begin{equation}\label{example}
    E=\frac{-v^2\sin(\theta)\cos(\theta)}{\sqrt{[1-v^2\cos^2(\theta)][1-v^2\sin^2(\theta)]}}.
\end{equation}
Of course, the same result is obtained with the use of  (\ref{m_sg}). We see that the expectation value tends to zero when $v$ is small, as it should be. Spin and momentum can then be treated independently, and in a non-relativistic scenario the expectation value of the measurement is the scalar product of unitary vectors in the direction of the apparatuses magnetic fields. However, if the particle has a relativistic speed this is not the case anymore. In figure \ref{fig1} we plot the expectation value for $\theta=\pi/4$ and $v$ between 0 and 1. When $v\rightarrow 1$, the expectation value tends to $-1$. This, of course, is surprising and would not be expected from a simple-minded analysis in which spin and momentum are assumed to behave independently. In fact, this example shows why, under general circumstances, we cannot treat the spin and the momentum of a relativistic particle as independent variables. If we trace out the particle momentum, we cannot correctly predict the expectation value of the spin measurement, even if the particle is in a momentum eigenstate.

\begin{figure}\begin{center}
  \includegraphics[width=10cm]{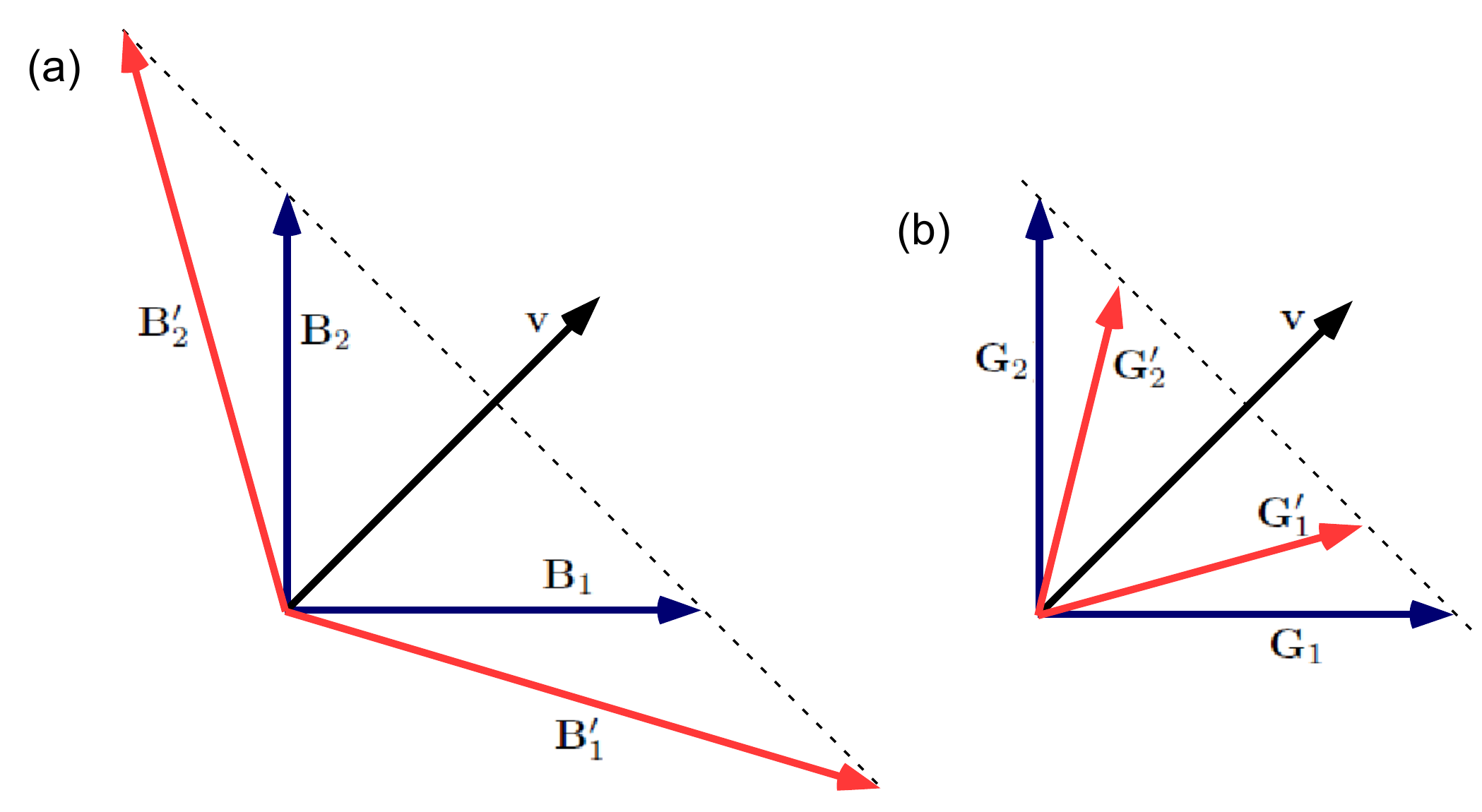}\\
  \caption{(a) Magnetic fields $\mb{B}_1$ and $\mb{B}_2$ in the laboratory frame are transformed to magnetic fields $\mb{B}_1'$ and $\mb{B}_2'$ in the rest frame of a particle that moves with relativistic velocity $\mb{v}$. (b) Fields $\mb{G}_1$ and $\mb{G}_2$ that are part of 4-vectors $(0,\mb{G}_1)$ and $(0,\mb{G}_2)$ in the laboratory frame are transformed to fields $\mb{G}_1'$ and $\mb{G}_2'$ in the rest frame of a particle that moves with relativistic velocity $\mb{v}$.
}\label{fig2}
 \end{center}\end{figure}

If we treat the same experiment using the Pauli-Lubanski formalism, hypothetically assuming that spin couples to 4-vectors $(0,\mb{G})$ in the laboratory frame, we find that the expectation value of the measurement of the second apparatus is $-E$ from  (\ref{example}), with the opposite sign in relation to the Stern-Gerlach case. We obtain completely different results with the two formalisms when the particle has a relativistic velocity. The reason for the difference is illustrated in figure \ref{fig2} for $\theta=\pi/4$. If we have Stern-Gerlach apparatuses with magnetic fields $\mb{B}_1$ and $\mb{B}_2$ in the laboratory frame, in the particle rest frame the fields $\mb{B}_1'$ and $\mb{B}_2'$ will have the same components parallel to the particle velocity, but the orthogonal components will be multiplied by $\gamma_v$. So the fields $\mb{B}_1'$ and $\mb{B}_2'$ tend to point in the opposite direction as the particle velocity approaches the speed of light, as depicted in figure \ref{fig2}-(a). However, if spin couples to 4-vectors $(0,\mb{G}_1)$ and $(0,\mb{G}_2)$ in the laboratory frame, in the particle rest frame the fields $\mb{G}_1'$ and $\mb{G}_2'$ will have the same components parallel to the particle velocity, but the orthogonal components will be divided by $\gamma_v$. So the fields $\mb{G}_1'$ and $\mb{G}_2'$ tend to point in the same direction as the particle velocity approaches the speed of light, as depicted in figure \ref{fig2}-(b).

\section{Conclusions}\label{sec:conc}

To summarize, we have provided a physical interpretation of the Wigner rotations in the quantum information context, that result from the fact that different observers compute different quantization axes for spin measurements. Based on that, we computed the expectation values of spin measurements made on relativistic spin-1/2 particles and concluded that it is not possible to measure the particle spin independently from its momentum, such that a momentum-spin partition of the system is actually completely meaningless. It is important to stress that our results for the expectation values of spin measurements should be valid for any measuring procedure in which spin couples to electromagnetic fields in the measuring apparatus, not only for the Stern-Gerlach apparatus. We also discussed that the spin operators must transform in the same way as the physical quantity to which they couple in the measuring apparatus in order to guarantee that all observers in inertial reference frames compute the same expectation value for the measurements. We presented an experimental proposal for the verification of our predictions, that we believe may be tested using an apparatus similar to the one in Sakai \textit{et al.} experiments \cite{sakai06}. 

The use of the spin of massive relativistic particles as the carriers of information in quantum information protocols may soon become a reality, and the preset work sets a formalism to be used for such tasks. For instance, our treatment predicts important effects for the violation of Bell's inequalities with entangled relativistic particles \cite{saldanha11}.

\ack

The authors acknowledge Daniel R. Terno and Nicolai Friis for fruitful discussions. P.L.S. was supported by the Brazilian agencies CAPES, CNPq and FACEPE. V.V. acknowledges financial support from the National Research Foundation and Ministry of Education in Singapore and the support of Wolfson College Oxford.

\appendix
\setcounter{section}{1}
\section*{Appendix}

We reproduce here the treatment of Weinberg \cite{weinberg} to describe the Wigner rotations. We can define states of 4-momentum $p=(p^0,\mb{p})$ for a particle with mass $m$ as
\begin{equation}\label{psi_p}
	|{p,\phi}\rangle\equiv\sqrt{\frac{k^0}{p^0}}U(L(p))|{k,\phi}\rangle,
\end{equation}
where $\phi$ denotes the spin state and $k=(m,0)$ is a standard 4-momentum, chosen to be in the particle rest frame. We are using a system of units in which the speed of light in vacuum is $c=1$. We have $p^\mu=\sum_\nu L^\mu_{\;\;\nu}(p)k^\nu$, or $p=L(p)k$, $L(p)$ being a standard boost, and $U(L(p))$ is the unitary transformation that transforms the state $|{k,\phi}\rangle$ in the state $|{p,\phi}\rangle$, $\sqrt{{k^0}/{p^0}}$ being a normalization factor.

If we make a change of the reference frame represented by a homogeneous Lorentz transformation $\Lambda$, the state in the new frame will be
\begin{eqnarray}\label{a2}\nonumber
	U(\Lambda)|{p,\phi}\rangle&=&\sqrt{\frac{k^0}{p^0}}U(\Lambda L(p))|{k,\phi}\rangle\\&=&\sqrt{\frac{k^0}{p^0}}U(L(\Lambda p)) 
	U(L^{-1}(\Lambda p)\Lambda L(p)) |{k,\phi}\rangle.
\end{eqnarray}
The transformation $W(\Lambda,p)\equiv L^{-1}(\Lambda p)\Lambda L(p)$, where $L^{-1}$ represents the inverse of $L$, leaves $k$ invariant, being a rotation denoted Wigner rotation, and we can write
\begin{equation}\label{a3}
	U(W)|{k,\phi}\rangle=\sum_{\phi'}D_{\phi,\phi'}(W)|{k,\phi'}\rangle,
\end{equation}
the matrix $D_{\phi,\phi'}(W)$ realizing the Wigner rotation on the particle spin. Substituting (\ref{a3}) and (\ref{psi_p}) in (\ref{a2}) we arrive at (\ref{wigner}).


\end{document}